\documentclass[pre,showpacs,amssymb,amsfonts,amsmath,floatfix,twocolumn]{revtex4-1}

\usepackage{natbib}
\usepackage{hyperref}
\usepackage{graphicx}% Include figure files
\usepackage{epsfig}% Include figure files
\usepackage{dcolumn}% Align table columns on decimal point
\usepackage{bm}% bold math
\usepackage{times}

\begin{document}
\title{Numerical estimate of the Kardar Parisi Zhang universality class in (2 + 1) dimensions}
%\title{Disproving Kim-Kosterlitz conjecture in (2+1) Dimensional Kardar-Parisi-Zhang universality class}

\author{Andrea Pagnani}
\affiliation{Department of Applied Science and Technology (DISAT), Politecnico di Torino, Corso Duca degli Abruzzi 24, I-10129 Torino, Italy}
\affiliation{Human Genetics Foundation (HuGeF), Via Nizza 52, I-10126, Turin, Italy}
\author{Giorgio Parisi}
\affiliation{Dipartimento di Fisica, INFN - Sezione di Roma 1, CNR-IPCF UOS Roma, Universit\`a ``La Sapienza'', P.le Aldo Moro 2, I-00185 Roma, Italy}

\begin{abstract}
We study the Restricted Solid on Solid model for surface growth in
spatial dimension $d=2$ by means of a {\em multi-surface coding}
technique that allows to produce a large number of samples of samples
in the stationary regime in a reasonable computational time. Thanks
to: (i) a careful finite-size scaling analysis of the critical
exponents, (ii) the accurate estimate of the first three moments of
the height fluctuations, we can quantify the wandering exponent with
unprecedented precision: $\chi_{d=2} = 0.3869(4)$. This figure is
incompatible with the long-standing conjecture due to Kim and
Koesterlitz that hypothesized $\chi_{d=2}=2/5$.
\end{abstract}

\maketitle 

The Kardar-Parisi-Zhang (KPZ) equation \cite{KPZ86} is one of the
simplest and most studied model of out-of-equilibrium surface
growth. The equation describes the time evolution of the height
$h({\mathbf r}, t)$ of an interface above a $d-$dimensional substrate:
\begin{equation}
\label{eq:KPZ}
\partial_t h({\mathbf r}, t) = \nu {\vec \nabla}^2 h({\mathbf r}, t) +
\frac \lambda 2 | {\vec \nabla}h({\mathbf r}, t)|^2 +\eta({\mathbf r}, t) \, ,
\end{equation}
where $\nu$ is the diffusion coefficient, $\lambda$ is the strength of
the non-linear growth rate, and $\eta({\mathbf r},t)$ is a Gaussian
white noise of amplitude $D$:
\begin{equation}
\label{eq:noise}
\langle \eta \rangle = 0 \,\,\,\,\,\,,\,\,\,\,\,\,
\langle \eta({\mathbf r},t) \eta({\mathbf r}',t') \rangle 
= 2D \delta^d({\mathbf r} - {\mathbf r}')
\delta(t-t')\,\,. 
\end{equation}
The universality class induced by Eq.~(\ref{eq:KPZ}) is defined in
terms of the scaling properties of the height fluctuations $w_2(L,t)
= \langle ( h ({\mathbf r}, t) - \langle h ({\mathbf r}, t)\rangle)^2
\rangle$. As a function of the system size $L$ it is believed that
$w_2(L,t) \sim L^{2 \chi}f(t/L^z)$, where the scaling function is such
that $f(x) \rightarrow \mathrm{const}$ for $x \rightarrow \infty$ and
$f(x) \sim x^{2\chi/z}$ for $x \rightarrow 0$. The peculiar behavior
of $f$ imply that $w_2(L,t) \sim L^{2\chi}$ for $t \gg L^z$ and
$w_2(L,t)\sim t^{2 \chi/z}$ for $t \ll L^z$. Due to an infinitesimal
tilt symmetry of Eq.~(\ref{eq:KPZ}) ($h\rightarrow h+{\mathbf r
}\cdot{\boldsymbol \epsilon}$, ${\mathbf r} \rightarrow {\mathbf r} -
\lambda t {\boldsymbol \epsilon} $), the two critical exponents are
related by the scaling relation $\chi + z = 2$, which is believed to
be valid at any dimension $d$ \cite{FamilyVicsekBook, *BarabasiBook}.

After decades of intense work in the field, the determination of the
two critical exponents $\chi,z$ is known rigorously only for $d=1$
where a fluctuation-dissipation relation leads to the exact result
$\chi = 1/2$, $z=3/2$. At any $d>1$ the quest for the quantification
of the critical exponents is still an open challenge. In particular,
in $d=2$ there is a long-standing conjecture dating back to the
seminal paper of Kim and Kosterlitz (KK) \cite{KimKosterlitz1989},
which proposes $\chi = 2/5,\,z = 8/5$. Such a conjecture has been
supported by a Flory type scaling argument in
\cite{Hentschel_FamilyPRL1991} and later by a field theoretical
operator product expansion in \cite{LaessigPRL1998}. More recently, a
nonperturbative renormalization group approximation reported a value
of $\chi = 0.33$ \cite{Canet_PRL2010,Canet_PRE2011} which, as we will
see in the following, is too small compared with our precise
measurements. From a numerical point of view the KK conjecture has
been put under scrutiny many times in the past
\cite{Chin_denNijs1999,Kondev_Henley_Salinas2000,NOIKPZ2001,Aarao-Reis2001,Aarao-ReisMiranda2008,Odor2009,
  KellingOrdor2011, Halpin_Healy_PRL2012} using different models
belonging to the KPZ universality class and different simulation
techniques.

In table Table~\ref{tab:confronti} we resume, at the best
of our knowledge, the current state of the art with respect to the
numerical check of the KK conjecture: although the statistical
uncertainty (when presented in the reference paper) is often too large
to exclude the validity of the KK conjecture, yet it is somehow clear
that all results fall somehow below the predicted rational
guess. Another common feature reported in the previously cited
bibliography, is that finite-size scaling corrections to the exponent
estimate seem to be particularly relevant, although very few work so
far have implemented a systematic procedure to take them into account.

\begin{center}
\begin{table}[htb]
\begin{tabular}{|*{4}{c|}}
\hline
Reference & $\chi$ & Model & Annotation\\
\hline
\hline
\cite{Forrest_Tang_PRL1990} & 0.385(5) & HSM & MC\\
\hline
\cite{Chin_denNijs1999}& 0.38(1) & BCSOS & FSS\\
\hline
\cite{Kondev_Henley_Salinas2000} & 0.38(8)& RSOS & non-linear
measures\\  
\hline
\cite{NOIKPZ2001} & 0.393(3) & RSOS & multispin-coding and FSS \\
\hline
\cite{Aarao-Reis2001} & 0.366, 0.363 & BD & MC and FSS\\
\hline
\cite{Aarao-ReisMiranda2008} & $0.38 \leq \chi \leq 0.40$& KPZ &
direct integration\\
\hline
\cite{Odor2009} & 0.377(15) & DLC & MC and FSS\\
\hline
\cite{KellingOrdor2011} & 0.393(4) & DLC &\begin{minipage}{1.in}
  bit-coded MC on GPUs and FSS \end{minipage}\\
\hline
\cite{Halpin_Healy_PRL2012} & 0.388 & KPZ & Eulerian integration\\
\hline
\cite{Halpin_Healy_PRL2012} & 0.385(4) & DPRM & transfer matrix method\\
\hline
\end{tabular}
\protect\caption{In this table we display the estimates in different
  previous work for the exponent $\chi$ (with the uncertainty when
  available), the model used (HSM = hypercubic stacking model, BCSOS =
  body centered solid-on-solid, KPZ is the direct integration of
  Eq.~(\ref{eq:KPZ}), BD = ballistic deposition, DLC = dimer lattice
  gas is a mapping to a discrete model described in details in
  \cite{Odor2009}, DPRM = direct polymer in random medium), and the
  integration method used (MC = monte-carlo, FSS = finite-size
  scaling). The estimate and uncertainty of the last row
  is obtained by averaging over two results obtained on simple cubic
  lattices with gaussian and uniform bond respectively.}
\label{tab:confronti}
\end{table}
\end{center}

\begin{figure}[ht]
\begin{center}
\includegraphics[width=\columnwidth]{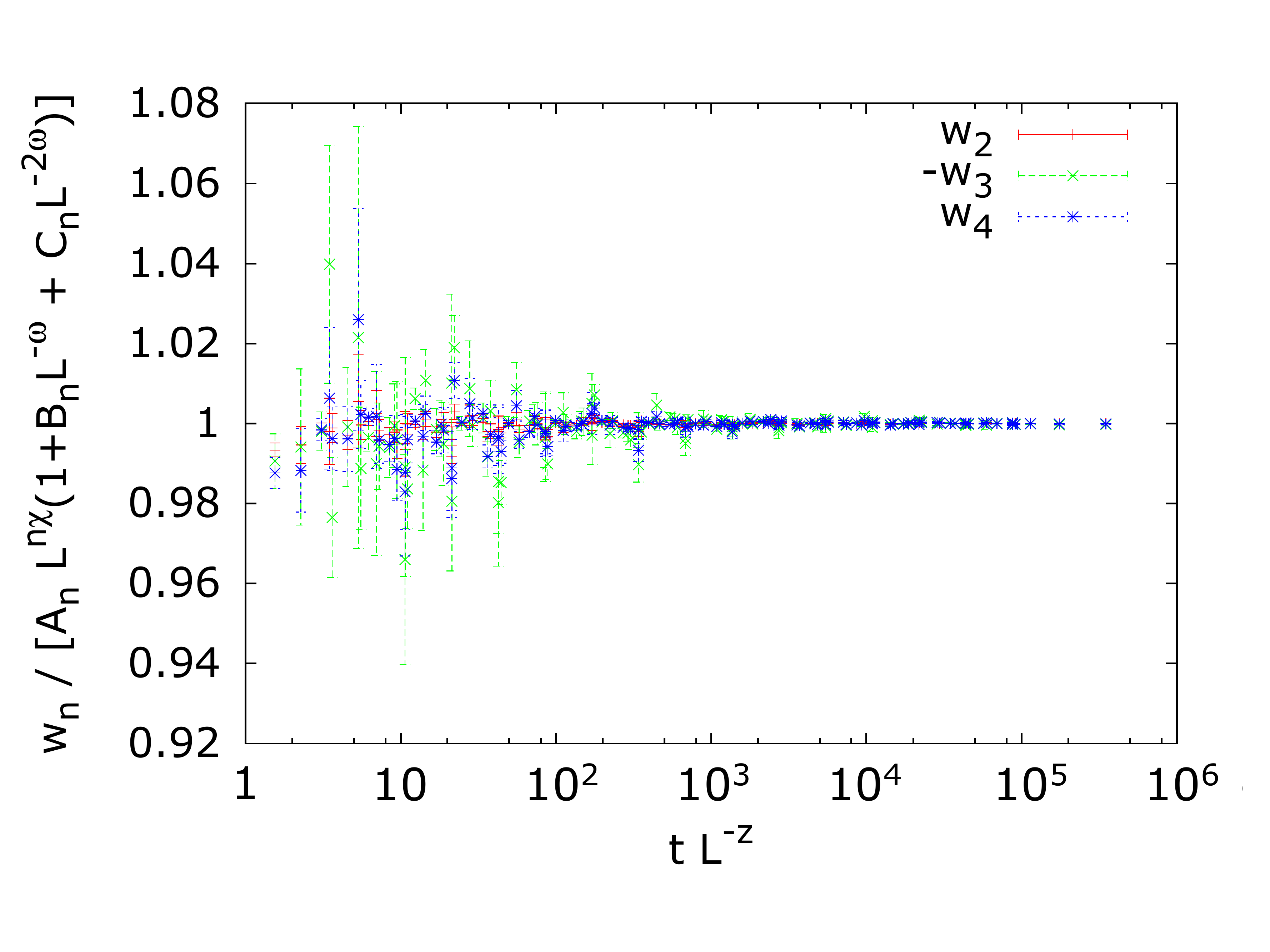}
\end{center}
\caption{\label{fig:scaling} (Color online) Scaling plot of the
  rescaled second moment $w_2/(A_2L^{2\chi}(1+B_2L^{-\omega}))$
  vs. the rescaled time $t/L^z$.}
\end{figure}

\begin{figure}[ht]
\begin{center}
\includegraphics[width=\columnwidth]{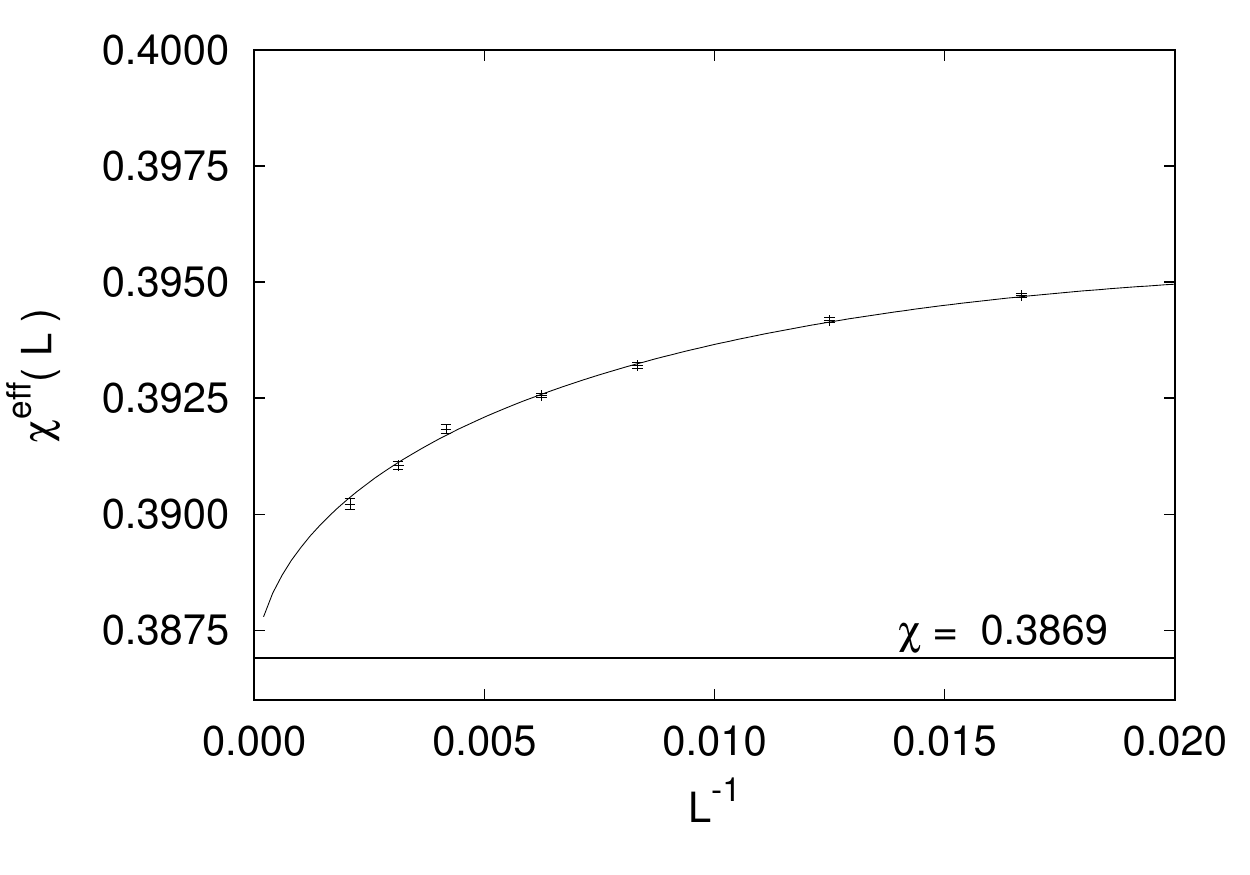}
\end{center}
\caption{\label{fig:localchi} Local slope of $w_2$ is displayed as a
  function of $L^{-1}$. Dots with error bars are values obtained by
  simulations, while the line is the 11-parameters best-fit reported
  in table~\ref{tab:fitpara}. The solid horizontal line is at
  $\chi=0.3869$, i.e. the best-fit prediction for the wandering
  exponent. The highest tick on the y-axis is 0.4 which is the KK
  conjecture.}
\end{figure}

Here, we will investigate the steady state scaling regime $t\gg L^z$
of a Restricted Solid on Solid (RSOS) model for lattice size volumes
up to $V=480^2$ of a very large number of surface samples to reduce as
much as possible the statistical error in the estimate of the critical
exponent. The RSOS model can be simulated in the following way: at any
time $t$ we randomly select a site $i$ on the $d-$dimensional lattice
and we let the surface height $h_i$ at that point to grow of a unit
$h_i(t+1) = h_i(t)+1$ only if $\max_{j \in \partial i} | h_i(t) -
h_j(t)| \leq 1$, where with $\partial_i$ is the set of 4 nearest
neighbors of $i$ in $d=2$ assuming periodic boundary conditions. We
used an improved multi-spin coding algorithm which has already been
described in detail elsewhere \cite{NoiPRE2013}.  We simulated
$2-$dimensional lattices of volume $V = L^2$ for lattices of linear
size $L=26,30,40,60,80,120,160,240,320,480$. A summary of our
simulations is provided in Table~\ref{tab:simulpara}.
\begin{center}
\begin{table}[htb]
\begin{tabular}{|*{3}{c|}}
\hline
 $L$ & $\log_2$(\#sweeps)  & \#samples \\
\hline
\hline
26 & 24 & 96  \\
\hline
30 & 24 & 1140   \\
\hline
40 & 27 &  30  \\
\hline
60 & 26 & 280  \\
\hline
80 & 27 & 60  \\
\hline
120 & 27 & 312  \\
\hline
160 & 27 & 156  \\
\hline
240 & 27 & 305  \\
\hline
320 & 27 &  492 \\
\hline
480 & 25 & 687  \\
\hline
\end{tabular}
\protect\caption{In this table we display the lattice linear size $L$,
  the base 2 logarithm of the number of montecarlo sweeps (full
  lattice updates), and the number of samples produced in our
  simulations.}
\label{tab:simulpara}
\end{table}
\end{center}
The numerical strategy adopted here is to achieve a fair statistical
sampling of the asymptotic regime $t > L^z$. At any time $t$ and for
each sample we measure the first three connected moments $w_n(L,t) =
\sum_{i=1}^V (h_i(t) - \langle h(t)\rangle)^n/V$, where $\langle h(t)
\rangle = \sum_{i=1}^V h_i(t)/V$, and $n=2,3,4$. Eventually, we define
our asymptotic (in time) estimate as:
\begin{equation}
w_n(L) = \frac1{T_0-T_1+1} \sum_{t=T_1}^{T_0}w_n(L,t )\,\,\,\,.
\end{equation}
In this way in practice we just consider the second half of the
simulation being able at the same time to judge how deep in the
asymptotic regime ($t \gg L^z$) our simulations are:
Fig.~\ref{fig:scaling} shows clearly that our data for all lattice
size produce a fair sampling of the steady state regime.
We already mentioned how relevant finite-size scaling correction to
the critical exponent are in 2 dimensional KPZ. To keep under control
in a reliable way the size dependence of the scaling we opted to fit
simultaneously the first 3 moments $w_{2,3,4}$, which at the leading
order, scale as $L^{n\chi}$ with $n=1,2,3$. The first order
corrections to the scaling are encoded in the exponent $\omega$ in the
following way \cite{NOIKPZ2001,NoiPRE2013}:
\begin{eqnarray}
\label{eq:fit-I}
w_2  &=& A_2 L^{2\chi}( 1 + B_2 L^{-\omega})\nonumber\\
w_3  &=& S A_2^{3/2} L^{3\chi}( 1 + B_3 L^{-\omega})\\
w_4  &=& K A_2^2 L^{4\chi}( 1 + B_4 L^{-\omega})\nonumber\,\,\,\,\,, 
\end{eqnarray}
As we will see in the following, finite-size scaling corrections turn
out to be particularly severe, so we analyzed our data using the
following second order fitting scheme:
\begin{eqnarray}
\label{eq:fit-II}
w_2  &=& A_2 L^{2\chi}( 1 + B_2 L^{-\omega} + C_2L^{-2\omega})\nonumber\\
w_3  &=& S A_2^{3/2} L^{3\chi}( 1 + B_3 L^{-\omega} + C_3L^{-2\omega})\\
w_4  &=& K A_2^2 L^{4\chi}( 1 + B_4 L^{-\omega} + C_4L^{-2\omega})\nonumber\,\,\,\,\,.
\end{eqnarray}
The relevance of the finite size scaling corrections is best
appreciated from Table~\ref{tab:fitcomp}, where we display, as a
function of the minimal linear lattice size, the outcome of the
fit. As far as the scheme proposed in Eq.~(\ref{eq:fit-I}) is
concerned, we see clearly how the larger is the lattice size, the
lower is the best-fit value for $\chi$. The variance of the reduced
$\chi$-square, although decreasing sensibly in the size interval
considered, due to the extreme precision in our estimation of the
moments, remains too large. The scenario becomes even more
satisfactory with the second fitting scheme defined in
Eq.~(\ref{eq:fit-II}) where, upon increasing the minimal lattice size,
the resulting best-fit values remain remarkably stable with a reduced
$\chi$-square around 1. For these reasons we choose as best fitting
scheme Eq.~(\ref{eq:fit-II}) using as minimal linear size $L=26$. Our
final estimate yields $\mathbf \chi = 0.3869(4)$ and $\omega =
0.56(5)$, and the best-fit values for the two fitting schemes are
reported in Table~\ref{tab:fitpara}. The values $\chi = 0.33$,
$\omega=0.7$ reported in \cite{Canet_PRL2010,Canet_PRE2011} are
still far from our numerical estimate.
\begin{center}
\begin{table}[htb]
\begin{tabular}{|*{5}{c|}}
\hline 
\multicolumn{5}{|c|}{FIT I Eq.~(\ref{eq:fit-I})}\\
\hline
Starting $L$ & $\chi$ & $\omega $ & $\sqrt{\mathrm{WSSR/NDF}}$ & NDF \\
\hline
26 &0.3904(4) & 0.9(1)& 8.53 & 22 \\
\hline
30 & 0.3903(4) & 0.9(1)& 8.41 & 19 \\
\hline
40 & 0.3898(5)& 0.9(1) & 5.36 & 16 \\
\hline
60 & 0.3893(6) &0.9(2) & 3.22 & 13\\
\hline
80 & 0.3892(7) &0.9(3) & 2.40 & 10\\
\hline
\hline
\multicolumn{5}{|c|}{FIT II Eq.~(\ref{eq:fit-II})}\\
\hline
26 &0.3869(4) & 0.57(5)& 1.2085 & 19\\
\hline
30 & 0.3866(6) & 0.53(6) & 1.27627 & 16\\
\hline 
40& 0.3868(7) & 0.5(1) & 1.22814 & 13 \\
\hline
60 & 0.383(4) & 0.3(2) & 1.09807 & 10\\
\hline
%% \multicolumn{5}{|c|}{FIT II Eq.~(\ref{eq:fit-II})}\\
%% \hline
%% 26 &0.3869(4) & 0.56(5)& 1.2084 & 19\\
%% \hline
%% 30 & 0.3866(6) & 0.53(6) & 1.27618 & 16\\
%% \hline 
%% 40& 0.3868(7) & 0.5(1) & 1.22795 & 13 \\
%% \hline
%% 60 & 0.383(4) & 0.3(2) & 1.09794 & 10\\
%% \hline
\end{tabular}
\protect\caption{In this table we display the lattice linear size the
  minimal lattice size from we start fitting the data, the best fit
  estimates of the exponents $\chi$ and $\omega$, the variance of the
  reduced $\chi$-square per degree of freedom
  ($\sqrt{\mathrm{WSSR/NDF}}$), and the number of degrees of freedom.}
\label{tab:fitcomp}
\end{table}
\end{center}
To appreciate more clearly the finite-size effects on $\chi$, we
evaluate the effective exponent $\chi_2^\mathrm{eff}$ as the discrete
logarithmic derivative of Eqs.~(\ref{eq:fit-II}), which in our case
reads:
\begin{equation}
\label{eq:chieff}
\chi_2^\mathrm{eff} (L) = \frac{\log(\frac{w_2(L)}{w_2(L')})}
{2 \log(\frac L{L'})}\,\,\,\,\,\,,
\end{equation}
where $L/L'= 2$. In Fig.~\ref{fig:localchi} we display
$\chi_2^\mathrm{eff}$ as a function of $L^{-1}$ and we superpose to
the data points the best-fit estimate for $\chi$. Note also how the
effective exponent, upon increasing the lattice size, departs
substantially from the KK conjecture value $\chi_{KK}=2/5=0.4$ which
in Fig.~\ref{fig:localchi} coincides with the upper extremal y-axis tick.

\begin{figure}[ht]
\begin{center}
\includegraphics[width=\columnwidth]{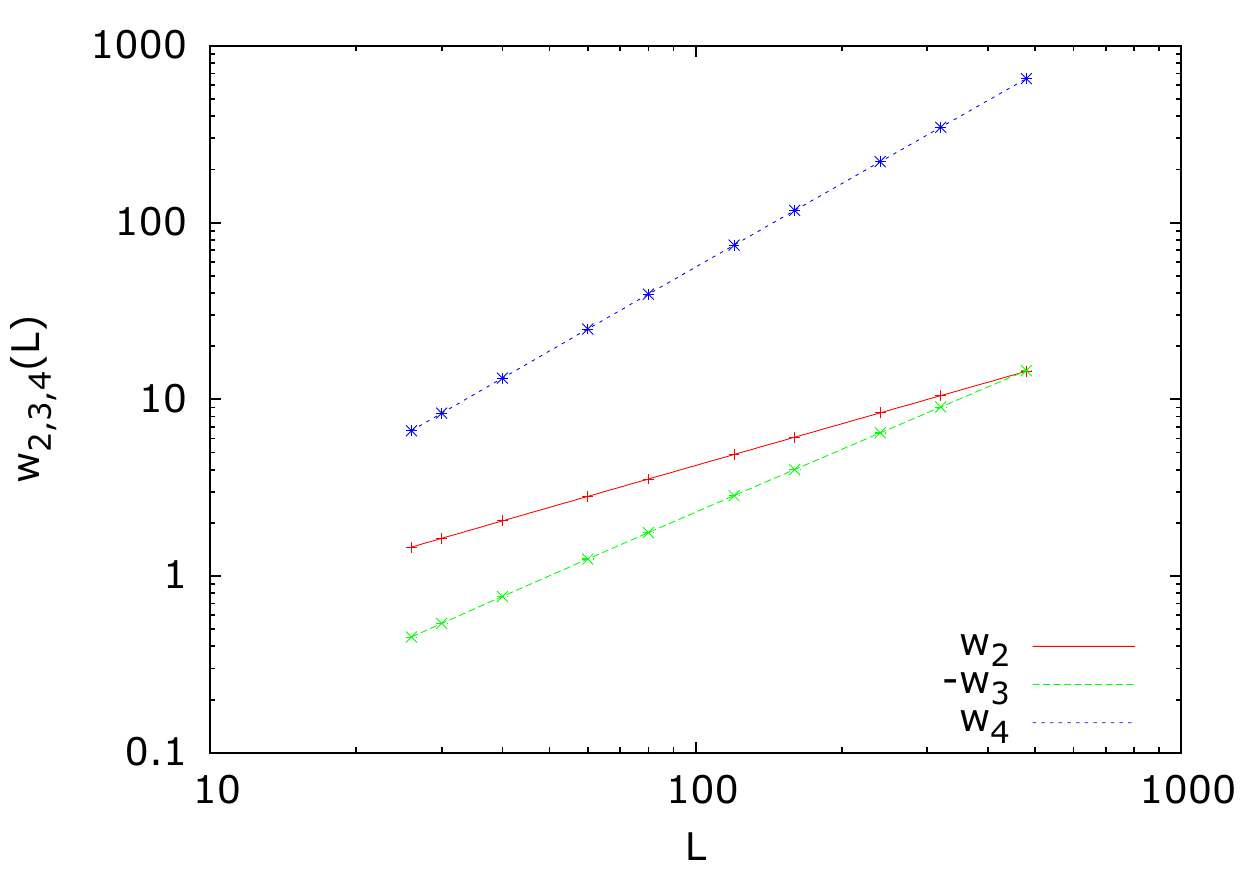}
\end{center}
\caption{\label{fig:scal3cum} (Color online) The quantities
  $w_{2},\ -w_{3}, \ w_{4}$ obtained by the best-fit value of
  Eq.~(\ref{eq:fit-II}) are displayed as a function of $L$ (lines) on
  double logarithmic scale. Dots with error bars are values obtained
  by simulations. Note that all $w_2$ values are larger than one even
  for the smaller lattice sizes.}
\end{figure}

\begin{figure}[ht]
\begin{center}
\includegraphics[width=\columnwidth]{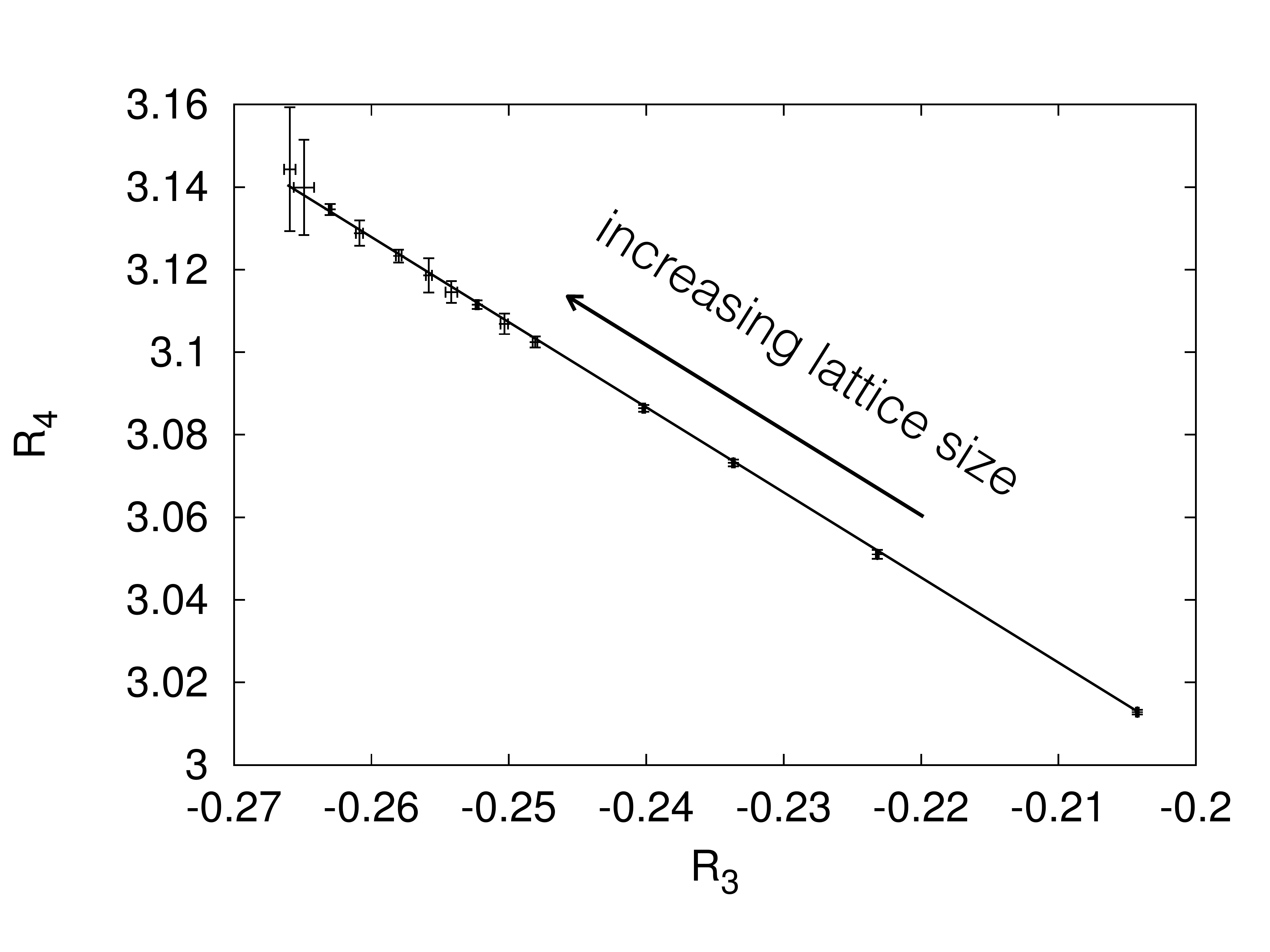}
\end{center}
\caption{\label{fig:ratios} Scatter plot of the ratio of the cumulant
  $R_4 = w_4/w_2^2$ vs. $R_3 = w_3/w_2^{3/2}$. Continuous line is the
  a linear fit to the data. Note that the lowest tick in y-axis should
  represent the value of the $R_4$ ratio for a normal distribution.}
\end{figure}

A matter of concern, when studying numerically scaling related
properties of system at criticality, is the ability to define how
deeply inside the critical phase the system under study is. For a
discrete model such as RSOS, a practical way to check this property is
to compare to the typical size of the fluctuations (given by $w_2$)
with the lattice spacing, which in our model is equal to 1
\cite{ColaioriMoore2001}. In Fig.~\ref{fig:scal3cum} we display the
values of $w_{2,3,4}$ measured in our simulations as a function of the
linear size of the systems. We can easily convince ourselves that all
our simulation are characterized by typical fluctuations which are
larger than the lattice spacing. At odds with what happens in $d=1$,
where in the asymptotic regime the fluctuation of the surface are
known to be Gaussian, the moments of the distribution show a strong
departure from the $d=1$ case. This is best appreciated in terms of
the ratio of the cumulants $R_4 = w_4/w_3^2$ vs. $R_3 = w_3/w_2^{3/2}$
as shown in Fig.~\ref{fig:ratios}, where a scatter plot of $R_4$
vs. $R_3$ is shown (note that increasing lattice sizes are from right
to left). The linear scaling behavior of the plot was already observed
in \cite{Chin-DenNijs1999,NOIKPZ2001}, and here again is clearly
indicating a strong departure from a normal distributed fluctuation of
the surface, as $R_4^{\mathrm{gauss}} = 3$.

The numerical technique we developed \cite{NoiPRE2013}, allowed us to
run very accurate numerical simulations of the RSOS model in $d=2$. We
have been able to estimate with an unprecedented accuracy the critical
exponent $\chi = 0.3869(4)$ in a reasonable amount of computational
time. The typical fluctuations length-scale of our simulations and our
careful finite-size scaling analysis clearly indicate that: (i) the
system reached a controlled scaling regime, (ii) the measured scaling
exponents are reliable and not affected by a pre-asymptotic cross-over
regime, (iii) the distribution of the fluctuations is non-Gaussian.  A
shrewd use of the simultaneous fit of the three cumulants as a
function of the lattice size, we are finally able to disprove the KK
conjecture that the value of the exponent $\chi = 2/5$, a figure that,
given the small statistical uncertainty of our estimate, lays at more
than 32 standard deviations away from our prediction.
\newline
\newline
We are deeply grateful to Timothy Halpin-Healy for many interesting
discussions regarding our work. GP acknowledges the European Research
Council for the financial support provided through the ERC grant
agreement no. 247328.
\begin{widetext}
\begin{center}
\begin{table}[htb]
\begin{tabular}{|*{12}{l|}}
\hline
  & $\chi$ & $\omega$ & $A_2$ & $B_2$  & $C_2$ &$S$ &  $B_3$ & $C_3$
& $K$ & $B_4$ &$C_4$\\
\hline
\hline
FIT I  & 0.3893(6) & 0.8(2) & 0.118(1) & -0.4(2) & NA & -0.2669(4)
&-1.1(6)  &  NA & 3.146(2) & -0.9(5) & NA
\\
\hline
FIT II & {\bf 0.3869(4)} & {\bf 0.57(5)} &0.1226(1) & -0.37(2) & 0.6(2) &
-0.2657(4) & -0.46(7) & -1.0(1) & 3.145(1) & -0.73(6) & 1.0(3)
\\
\hline
\end{tabular}
\protect\caption{In this table we display the best fit values together
with their statistical error of the parameters defined in
Eqs.~(\ref{eq:fit-I} , \ref{eq:fit-II}).}
\label{tab:fitpara}
\end{table}
\end{center} 
\end{widetext}

%% \begin{widetext}
%% \begin{center}
%% \begin{table}[htb]
%% \begin{tabular}{|*{12}{l|}}
%% \hline
%%   & $\chi$ & $\omega$ & $A_2$ & $B_2$  & $C_2$ &$A_3$ &  $B_3$ & $C_3$
%% &$A_4$ & $B_4$ &$C_4$\\
%% \hline
%% \hline
%% FIT I  & 0.3893(6) & 0.8(2) & 0.118(1) & -0.4(2) & NA & 0.0108(2)
%% &-1.1(6)  &  NA & 0.0438(8) & -0.9(5) & NA
%% \\
%% \hline
%% FIT II & {\bf 0.3869(4)} & {\bf 0.56(5)} &0.1226(1) & -0.37(2) & 0.6(2) &
%% 0.0114(1) & -0.46(7) & 1.0(1) & 0.0473(7) & -0.73(6) & 1.0(3)
%% \\
%% \hline
%% \end{tabular}
%% \protect\caption{In this table we display the best fit values together
%% with their statistical error of the parameters defined in
%% Eqs.~(\ref{eq:fit-I} , \ref{eq:fit-II}).}
%% \label{tab:fitpara}
%% \end{table}
%% \end{center} 
%% \end{widetext}

%\begin{thebibliography}{100}
%\bibliography{kpz4d}

%\end{thebibliography}
\end{document}